# Reversal behavior of exchange-biased submicron dots

Zhi-Pan Li, [a) *] Oleg Petracic, [a), b)] Johannes Eisenmenger, [a), c)] and Ivan K Schuller [a)]

[a)] Physics Department, University of California, San Diego, La Jolla, CA 92093-0319

[b)] Angewandte Physik, Universität Duisburg-Essen, 47048 Duisburg, Germany

[c)] Abteilung Festkörperphysik, Universität Ulm, 89069 Ulm, Germany

**Abstract** - Nanostructured Fe dots were prepared on antiferromagnetic $FeF_2$ thin films and investigated by magneto-optical Kerr effect (MOKE). We studied the influence of dot sizes on the magnetic hysteresis and compared the result with both continuous thin film bilayers and nanostructured $Fe/FeF_2$ pillars. Hysteresis loops were measured at temperatures below and above (10 and 90 K, respectively) the Néel temperature of the antiferromagnet. A vortex state is found for dots of 300 nm diameter, where the exchange bias field is reduced compared to larger dot system and the continuous bilayer. Micromagnetic simulations including the interaction with the antiferromagnet show qualitatively similar behavior.
PACS numbers: 75.50.Ee, 75.60.Jk, 75.75.+a

---

[*] Corresponding author: Zhipan@physics.ucsd.edu

Submicron ferromagnetic (FM) dot arrays have attracted much attention recently.[1,2] This is driven by the technological interest in higher storage density[3], miniaturization of sensors and basic resarch in reduced dimensions. When the size of a magnetic system is reduced close to the order of 100 nm, interesting spin configurations can occur,[4] e.g. the single domain and vortex state. For dots of certain sizes, when the magnetic field is reduced from saturation, a vortex core nucleates at one edge, reversibly moves across the dot and annihilates on the other side. In this case, the reversible movement of the vortex core often manifests in the hysteresis loop as a straight line through the origin. This has been studied in detail both experimentally and using micromagnetic simulations.[1,4]

In this work we combine ferromagnetic dot arrays with an antiferromagnet (AF).[5-8] When such a system is cooled below the Néel temperature $T_N$ of the AF in a magnetic field $H_{CF}$, the interaction at the FM/AF interface gives rise to the exchange bias (EB) effect, revealed as a shift of the hysteresis loop along the field axis.[9] Many models have been proposed to explain EB.[10] Some of them address different magnetization reversal mechanisms and how they influence EB.[11] However, the detailed mechanism of EB is still unclear. Here, we modify the reversal process of a FM by shaping it into submicron dots to investigate its influence on the EB effect.

Fe(30 nm)/FeF$_2$(20 nm) bilayers capped with 4 nm Al were prepared on top of a single crystal MgO(100) substrate by e-beam evaporation. The FeF$_2$ grows as a twinned quasi-epitaxial film,[12] whereas the Fe layer is polycrystalline. Square arrays of circular Fe dots with diameter $d$ = 300 nm or 600 nm and center-to-center distance of $a = 2d$ were prepared by e-beam lithography and subsequent Ar$^+$-ion milling. By controlling the ion-milling time, two different types of systems were prepared from the same bilayer sample: in the type A sample, only the Fe layer was nanostructured, while in type B both Fe and $FeF_2$ were nanostructured (see Figure 1(a)). In both cases, a small area was kept unexposed to the ion beam to allow comparison of the dots and continuous film on the same sample. The dot arrays were imaged by atomic force microscopy as shown in Figure 1(b). The samples were initially cooled from 150 K to 10 K through the Néel temperature $T_N$ = 78.4 K of $FeF_2$ in an in-plane cooling field $H_{CF}$ = 5 kOe. Magnetic measurements were carried out using low-temperature magneto-optical Kerr effect (MOKE) both below and above the Néel temperature ($T$ = 10 and 90 K) to compare the reversal behavior when the AF is either ordered or in a paramagnetic state, respectively. The laser beam was focused to 50 μm, much larger than the dot size, thus measuring the average behavior of a large number of dots.

The results from the two types of dot arrays and the film are summarized in Table I. Figure 2 shows the measurement on sample A for Fe dots of 300 nm (a) and 600 nm (b) diameter respectively, at $T$ = 10 and 90 K. At $T$ = 90 K, the continuous Fe/$FeF_2$ film exhibits a square loop (upper inset of Figure 2(a)). The 300 nm dot array clearly shows that the two hysteresis branches almost join at zero field, which is characteristic of the vortex state.[1] The 600 nm dot array shows a sheared loop at 90 K without any vortex signature. The observed shearing of the loop is generally believed to come from shape anisotropy or the distribution of switching fields.[2] When the sample is cooled to $T$ = 10 K, the exchange bias of the continuous film manifests as a clear loop shift by $H_E$ = −97 Oe. While the 600 nm dots exhibits $H_E$ = −96 Oe similar to that of the continous film, the 300 nm dot array shows a smaller EB field $H_E$ = −55 Oe. The coercivity is enhanced upon biasing for both dot sizes. The collapse of the two hysteresis branches in the 300 nm dot leads to a smaller coercivity than the 600 nm dots. Moreover, patterning also leads to an increased coercivity compared with the unpatterned film possibly due to the increased importance of the shape anisotropy and pinning.

The type B samples (Figure 3) show much larger coercivities than type A with a similar trend in size and temperature as shown in Table I. This may be attributed to increased structural defects and pinning due to the high etching rate of the $FeF_2$ compared with Fe. Contrary to type A, the 300 nm type B dots do not show the vortex-characteristic narrowing in the hysteresis loop close to zero field, showing that the increased pinning modifies the reversal behavior with the vortex becoming pinned or inhibited. Moreover, the EB field becomes almost independent of the lateral size, i.e. -110 and -105 Oe for the 300 nm and 600 nm dot array respectively, which is comparable with the continuous film, $H_E$ = -97 Oe. Shaping the AF has little influence on the EB field possibly because the AF domain size in twinned $FeF_2$ is estimated to be close to the grain size of about 10 nm[13], which is much smaller than either dot dimension.

The above results imply several important features related to EB. First, the 300 nm dots of type A with a vortex-characteristic hysteresis exhibit a reduction of the EB field. For other cases, regardless of dot sizes and types, the unidirectional anisotropy shifts the hysteresis loops by $H_E \approx$ -100 Oe. Second, both dot types show larger coercivities at $T$ = 10 K than $T$ = 90 K. This observation is consistent with the coercivity enhancement commonly observed in EB systems.[10] It should be mentioned that several groups have reported a decrease of the EB field upon nanostructuring,[6,7,14] while the reverse situation was also observed.[8] A possible scenario is that different parts of in the thickness-diameter diagram have different size dependences of the EB. Additional experimental studies are needed to clarify this issue.

To understand the reversal behavior of the dots, we performed micromagnetic calculations.[15] First, the FM dots without the AF are simulated. The $d = 600$ nm and $t = 30$ nm Fe dots show a reversal through a multi-domain state with a similar magnetization curve as in our experiment (lower inset of Figure 2 (b)). For the 300 nm dots, the shape anisotropy dominates its behavior and a flux-closure vortex state is encountered (lower inset of Figure 2 (a)). This confirms our experimental observations that in type A dots a vortex state is observed in the 300 nm dots, but not in the 600 nm ones. The incomplete collapse in the experiment may arise from deviations from circular shape of the dots, roughness, structural variations from dot to dot and other imperfections.

To investigate the influence of the vortex state on the EB, we assume that 4% of randomly distributed, rigid, uncompensated AF interfacial spins are exchange coupled to the bottom layer of the FM[16,17] because of the very high anisotropy of $FeF_2$.[18] The interfacial coupling strength is taken to be the AF coupling in the $FeF_2$, $J_{FM/AF} = -0.45$ meV.[18] The results of these simulations are presented in Figure 4, where the 300 nm dot shows an EB field of -206 Oe, compared with -505 Oe for the continuous film. The same trend was found experimentally. Figure 5 shows the corresponding spin structure of the biased dot at different fields along the increasing hysteresis branch. There is virtually no difference in the reversal process compared to the unbiased case (see Fig. 2 (a) inset) except an overall EB shift. This means the AF pinning spins act as a uniform EB field, and the magnetization loop resembles that of the unbiased case. Moreover, the vortex core is no longer at the center of the dot at zero field, but shifted to one side in the direction perpendicular to the bias field. The reduction of the EB field for the vortex state can be understood since a flux closed state has only part of its spins pointing parallel to the net frozen interface moment of the AF, thus the total interfacial coupling energy is reduced. In other words, the exchange bias is associated with a term of the type $\vec{S}_{FM} \cdot \vec{S}_{AF}$ [19], which is reduced in the vortex state. On the other hand, the above simulation does not show any coercivity increase upon biasing as found in the experiment. Hence, the ansatz considering only few uncompensated unidirectional frozen AF moments is too simple. This might be related to the lack of a reversible component of AF interfacial spins leading to an additional contribution to the coercivity. This discrepancy can also come from variations among individual dots, e.g. distribution of defect pinning and exchange coupling. Further studies using other techniques like low temperature MFM can help make direct comparison with the micromagnetic simulation.

In conclusion, we have studied the reversal behavior of submicron Fe dots exchange biased to $FeF_2$ using low temperature MOKE. We varied the diameter of dots ($d$ = 300 and 600 nm) as well as the type of structuring, i.e. Fe dots on top of a continuous $FeF_2$ film (type A) or both Fe and $FeF_2$ patterned (type B), while the thickness was kept constant, $t_{FM}$ = 30 nm, $t_{AF}$ = 20 nm. We find that a vortex state leads to an EB field smaller than in all other cases. This result is consistent with micromagnetic simulations.

This work has been supported by AFOSR, DOE, the AvH Foundation, Cal-(IT)$^2$, and CNPq.

**TABLE I.** Coercivities $H_C$ at $T = 10$ K and 90 K, and exchange bias field $H_E$ at $T = 10$ K for dot arrays and continuous film determined from the inflection points of the hysteresis loops. The error is close to 5 Oe for the continuous film, 10 Oe for type A dots, and 20 Oe for type B dots.

| Type | $H_C$ (Oe) at $T = 90$ K | $H_C$ (Oe) at $T = 10$ K | $H_E$ (Oe) at $T = 10$ K |
|---|---|---|---|
| A, $d$=300 nm | 99 | 302 | -55 |
| A, $d$=600 nm | 395 | 546 | -96 |
| B, $d$=300 nm | 299 | 433 | -110 |
| B, $d$=600 nm | 768 | 875 | -105 |
| Continuous Film | 54 | 60 | -97 |

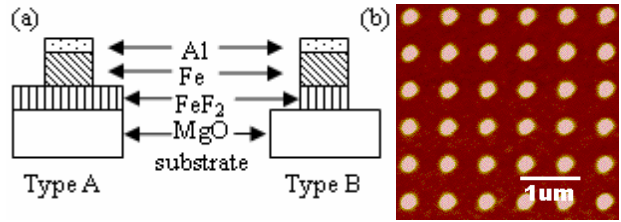

**Figure 1**. (a) Schematic of type A and B samples. (b) Atomic force microscopy image of a type B sample with dot diameter 300 nm. The array size is $80 \times 80$ $\mu m^2$.

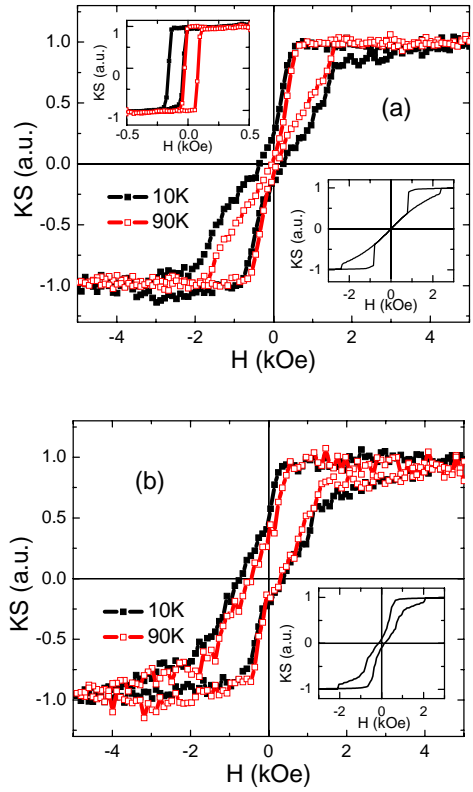

**Figure 2**. Kerr effect signal (KS) vs. magnetic field H from type A sample (Fe dots on FeF$_2$ film) with dot diameter 300 nm (a) and 600 nm (b) at $T$ = 10 K (solid squares) and 90 K (open squares). The upper inset of (a) shows the data on the continuous film of the same sample. The lower insets of (a) and (b) show corresponding data from micromagnetic calculations in the unbiased case.

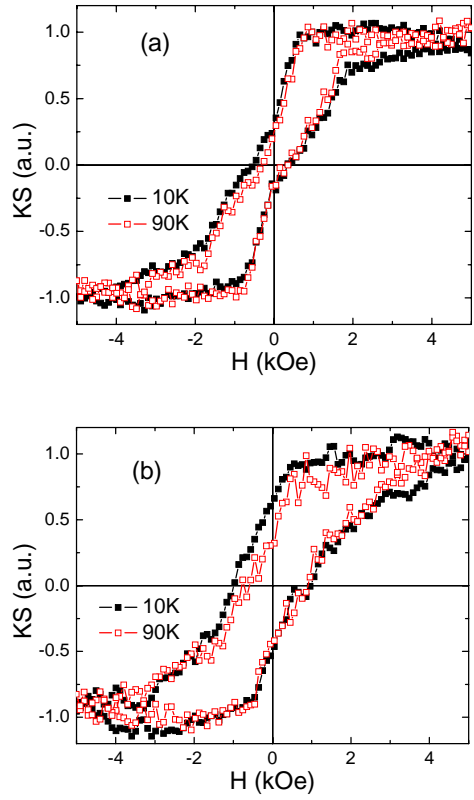

**Figure 3**. Kerr signal (KS) vs. magnetic field H from type B sample with dot diameter 300 nm (a) and 600 nm (b) at $T = 10$ K (solid squares) and 90 K (open squares).

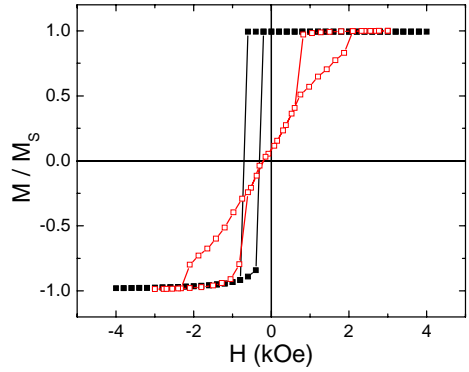

**FIGURE 4**. Micromagnetic simulations of a FM film (filled square) and a 300 nm type A dot (open square) subject to rigid AF uncompensated spins. The thickness in both cases is 30 nm.

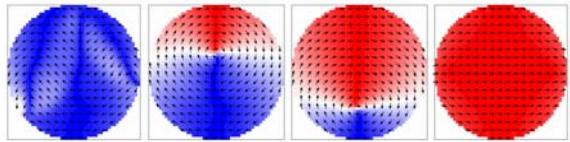

**FIGURE 5**. Spin configurations of a 300 nm type A dot in external fields of -825, -600, 975, and 2550 Oe along the increasing hysteresis branch from the micromagnetic simulation. The red, white and blue color codes refer to $M_x$ (horizontal direction) equal to 1, 0 and -1 respectively.